\documentclass[12pt,preprint]{aastex}

\newcommand\nH{n_\mathrm{H}}
\newcommand\percc{\textrm{cm}^{-3}}
\def\ee #1 {\times 10^{#1}}          
\def\ut #1 #2 { \, \textrm{#1}^{#2}} 
\def\un #1 { \, \textrm{#1}}          
\newcommand{\msol}{\,\textrm{M}_\sun}                
\newcommand{\kms}{km\,s$^{-1}$}

\begin{document}



\slugcomment{ApJL, in press}
\shorttitle{}
\shortauthors{}

\title{On the Formation of Compact Stellar Disks Around Sgr A*}
\author{Mark Wardle}
\affil{Department of Physics, Macquarie University, Sydney NSW 2109,
Australia}
\email{wardle@physics.mq.edu.au}
\author{Farhad Yusef-Zadeh}
\affil{Department of Physics \& Astronomy,
Northwestern University, Evanston, IL 60208, USA}
\email{zadeh@northwestern.edu}

\begin{abstract}
The recent identification of one or two sub-parsec disks of young,
massive stars orbiting the $\sim4\ee 6 \msol$ black hole Sgr A* has
prompted an ``in-situ'' scenario for star formation in disks of gas
formed from a cloud captured from the Galactic center environment.  To
date there has been no explanation given for the low angular momentum
of the disks relative to clouds passing close to the center.  Here we
show that the partial accretion of extended Galactic center clouds,
such as the 50 \kms\ giant molecular cloud, that temporarily engulf
Sgr A* during their passage through the central region of the Galaxy
provide a natural explanation for the angular momentum and surface
density of the the observed stellar disks.  The captured cloud
material is gravitationally unstable and forms stars as it
circularizes, potentially explaining the large eccentricity and range
of inclinations of the observed stellar orbits.  The application of
this idea to the formation of the circumnuclear ring is also
discussed.
\end{abstract}

\keywords{accretion, accretion disks --- Galaxy: center --- 
ISM: clouds --- stars: formation}
\section{Introduction}
\label{introduction} 

A high concentration of mass, almost certainly a $\sim 4\ee6 \msol$
black hole, is located at the very center of an evolved,
centrally-concentrated stellar population and coincident with the
bright compact radio source Sgr A* (e.g.\ Genzel et al.\ 2003;
Sch\"odel et al.  2003; Ghez et al.  2003; Reid \& Brunthaler 2004).
The discovery of a young cluster of massive stars (Krabbe et al.\
1991, 1995) in the hostile tidal environment within a parsec of Sgr A*
is surprising (e.g.\ Morris 1993); even more remarkable is the
discovery that the cluster consists of one or possibly two
counter-rotating, thick stellar disks with surface density profiles
scaling as the inverse square of the true distance from Sgr A* (Genzel
et al.\ 2003; Paumard et al.\ 2006), although the existence of the
second disk awaits independent confirmation (Lu et al.\ 2006).  The
inner and outer radii of the better-defined clockwise disk are
$\approx$0.03 and 0.3 pc, and the stellar ages are estimated to be
$\sim 6$\, Myr, with the total mass of stars amounting to $\sim
1.5\times 10^4\,\msol$  (Paumard et al.\ 2006).

There are two mechanisms favoured for the formation of a compact
stellar disk around Sgr A*.  In one, a cluster of massive stars
spirals into the central region because of tidal friction with the
evolved stellar population centered on Sgr A*, and is tidally
disrupted to form a stellar disk (Gerhard 2001; McMillan \& Portegies
Zwart 2003; Portegies Zwart, McMillan \& Gerhard 2003; Kim et al.\
2004; G\"urkan \& Rasio 2005).  The time scale for tidal friction
exceeds the stellar ages unless the cluster is extraordinarily massive
and compact (see Paumard et al.  2006), although recent work finds a
faster inspiral so that this possibility is still open (Fujii et al.\
2007).  Nevertheless, this mechanism tends to produce a far more
disordered stellar orbits than observed, as well as a population of
massive stars shed from the cluster that should extend beyond 0.3\,pc
from Sgr A*.  In the second, ``in-situ formation'' scenario, an
interstellar cloud is tidally disrupted and captured by Sgr A*,
settles into a gravitationally unstable disk, and forms the stars that
we see today (e.g.\ Levin \& Bolobordov 2003; Nayakshin \& Cuadra
2005).  In-situ star formation overcomes the time scale issue and more
naturally produces the gross kinematics of the observed disks.
However, star formation within a kinematically cold disk produces
stellar orbits that are less eccentric and more coplanar than
observed, even accounting for gravitational scattering of newly-formed
stars by other members of the new stellar population (Cuadra, Armitage
\& Alexander 2008).

Simulations of this scenario generally start with the evolution of an
initially gravitationally unstable disk or of a compact cloud in a
close orbit around Sgr A*, implicitly assuming that formation of the
stellar disk is initiated by the chance capture of an isolated,
compact, low-angular momentum gas cloud.  This, however, is unlikely
to be so because of the compactness of the stellar disk relative to
the scale of molecular clouds and cores.  While the kinetic energy of
a compact incoming cloud can be readily dissipated by shocks and
subsequently radiated away, the net angular momentum of the cloud
material remains unchanged.  Thus the initial angular momentum must be
very low indeed if the cloud is to circularize into a disk of radius
$\la$ 0.3\,pc.  Instead, as recently noted by Yusef-Zadeh \& Wardle
(2008), it is much more likely that the precursor disks are formed by
the partial capture of an extended molecular cloud that temporarily
engulfs Sgr A* on a passage through the Galactic center rather than
passing to one side of it.  Simulations by Mapelli et al.\ (2008) of
the capture of a cloud on an almost radial orbit suggests that star
formation during such an event will occur before the disk has fully
circularized and become dynamically cold, so this scenario may better
explain the observed kinematics.

Here we show that the observed stellar disk properties arise
naturally by the partial capture of an extended molecular cloud that
temporarily engulfs Sgr A* on a passage through the Galactic center
rather than passing to one side of it.  Cloud material passing on
opposite sides of Sgr A* have oppositely-directed angular
momenta, and gravitationally-induced collision of material downstream
of Sgr A* reduces its angular momentum, permitting the captured
material to settle into a compact configuration. The disks that form
from this process are typically highly gravitational unstable, and so
star formation can be expected to occur before the gas becomes
dynamically cold.  We also apply this formation scenario to the
circumnuclear ring of gas which encircles Sgr A* with a rotational
velocity of $\approx$100 \kms\ on a scale of 2-5 parsecs, and argue
that it is just settling down after a recent capture event and is on
the verge of forming stars.

\section{Cloud Capture by Sgr A*}

First we show that it is almost impossible for a cloud to be captured
by Sgr A* and circularize to form the progenitor of the observed
stellar disk without engulfing Sgr A* during the
encounter.  Consider an incoming cloud with velocity $v =
100\,v_{100}$\,\kms\ and impact parameter $b$ at infinity, and assume
that the cloud passes entirely to one side of Sgr A*\ as it begins the
process of circularization.  Tidal stretching and shocking convert the
cloud's bulk kinetic energy to heat, but there is no mechanism able to
reduce the mean angular momentum per unit mass $\sim b\,v$.  Therefore
the radius of the resulting disk $r_d$ satisfies
\begin{equation}
    r_d \left(\frac{GM}{r_d}\right)^{1/2} \approx b\,v\,,
    \label{eq:rd1}
\end{equation}
where $M = 4\ee 6 \msol $ is the mass of Sgr A*.  The observed stellar
disk size, $r_d\approx 0.3$\,pc, then implies that the impact
parameter $b\la 0.7\,v_{100}^{-1}$\,pc.  To avoid engulfing Sgr A*
during the capture, the cloud's radius must be much less than $b$,
implying an initial density $\nH \ga 10^7\percc$ if the resultant disk
is to have the mass $\sim 10^5\msol$ inferred for the progenitor of
the observed stellar disk.  This scenario therefore requires a compact
($\la 0.5$\,pc) and dense ($\ga 10^7\percc$) interstellar cloud to be
on a trajectory with impact parameter $\la 1$\,pc of Sgr A* -- a 
unique event.

Now consider the partial capture of clouds that engulf Sgr A* during
their passage through the inner few parsecs of the Galaxy.  The
capture is enhanced by the gravitational focussing of material passing
by Sgr A* and the subsequent collision of the gas just beyond Sgr A*,
in a manner analogous to Bondi--Hoyle--Lyttleton accretion (Bondi \&
Hoyle 1944).  Fluid elements passing on opposite sides of Sgr A* have
oppositely-directed orbital angular momenta, so that the collision
between them reduces their specific angular momentum.  The efficiency
of angular momentum cancellation depends on the density and velocity
inhomogeneities in the incoming material.  Velocity fluctuations are
negligible because the velocity dispersion within molecular clouds is
small compared to the highly supersonic bulk motion as clouds approach
Sgr A*.  Density inhomogneities in moleular clouds are large; however
their effect is mitigated because the collision-induced accretion rate
depends quadratically (rather than linearly) on the departure from
homogeneity (Davies \& Pringle 1980).  Numerical simulations have
confirmed that the cancellation in the face of asymmetries of order
unity is surprisingly efficient (Edgar 2004, and references therein).
One key difference from classic Bondi--Hoyle--Lyttleton accretion flow
is that the incoming gas is finite in extent.  This means that tidal
stretching of the incoming cloud may markedly change the outcome and
that the flow does not develop long-term average behaviour.

Simulations are needed to address the details of the circularization
process.  For now we characterise the uncertain capture dynamics using
two parameters: $\kappa$, the ratio of the captured mass to the
Hoyle--Lyttleton estimate (Hoyle \& Lyttelton 1939), and $\lambda$,
the fraction of the initial specific angular momentum retained by the
captured material.  These key parameters are sufficient to estimate
the gross features of the resulting disk of captured material: its
mass and size.  While in principle $\kappa$ and $\lambda$ lie between
0 and $\sim 1$, our expectation is that the relative ease with which
the gas can dissipate its bulk kinetic energy implies that
$\kappa\sim1$, wheras cancellation of angular momentum will be
imperfect because of the inhomogeneities and finite extent of the
incoming material, so that (perhaps) $\lambda\sim 0.3$.

Suppose then that an extended cloud with surface mass
density $\Sigma_\mathrm{cl}$ equivalent to a column density 
of hydrogen nuclei
$N_{24}\ee 24 $ H\,cm$^{-2}$
is passing through the
Galactic center with speed $v$.  
Cloud 
material with impact parameters less than about
\begin{equation}
    b_0 = \frac{2GM}{v^2} = 3.4\,v^{-2}_{100}\;\un pc\,,
    \label{eq:b0}
\end{equation}
from Sgr A* is captured, circularizes, and forms a disk of mass
\begin{equation}
    M_d = \pi \,\kappa\, b_0^2\, \Sigma_\mathrm{cl} 
        =  4.2\ee 5 \,\kappa\,N_{24}\,v^{-4}_{100}\;\, \msol\,.
    \label{eq:Md}
\end{equation}
The outer radius of the disk, $r_d$, has specific 
angular momentum $r_d\sqrt{GM/r_d}$; the disk material at this radius 
corresponds to the matter with the 
largest angular momentum prior to capture, with impact 
parameter $\sim b_0$.  The specific angular momentum of this material 
after circularization is $\lambda b_0 v$, so the outer radius of the 
disk is
\begin{equation}
    r_d = 2\lambda^2\, b_0 = 6.9 \, \lambda^2\, v^{-2}_{100} \; \un pc\,.
    \label{eq:rd}
\end{equation}
and the disk surface density is 
\begin{equation}
    \Sigma_d = \frac{M_d}{\pi r_d^2} 
             = \kappa\,\lambda^{-4}\,\Sigma_\mathrm{cl} 
             = 0.59 \, \kappa\,\lambda^{-4}\; \un g \ut cm -2  \,,
    \label{eq:Sigmad}
\end{equation}
Then Toomre's $Q$ is 
\begin{equation}
    Q = \frac{c_s\Omega}{\pi G\Sigma_d} 
      = 0.11\,T^{1/2}_{100}\,\frac{\lambda \,v_{100}^3 }{ \kappa N_{24}}
    \label{eq:Q}
\end{equation}
with gravitational instability possible when $Q<1$.  Here we have
scaled the expression to a gas temperature 100\,K, a reasonable lower
limit given the intense heating by the hot stars in the inner few
parsecs of the Galactic center.

Eqs (\ref{eq:Md})--(\ref{eq:Q}) show that the
gross properties of the resultant disk depend on only two independent
combinations of the four parameters $v_{100}$, $N_{24}$, $\kappa$ and
$\lambda$, namely $v_{100}/\lambda$ and $\kappa N_{24} / v_{100}^4$.
Note in particular that the temperature of the cloud does not affect the 
properties of the disk because the incident cloud material is highly 
supersonic.

In Fig.\ \ref{fig:parameter_space} we plot lines of constant disk 
mass, disk radius and Q in this two-dimensional parameter space, and
indicate the regions corresponding to the stellar disk around
Sgr A* and the circumnuclear ring (the latter is discussed in the next 
section).

The size of the stellar disk, 0.3\,pc, implies that $v_{100}/\lambda
\sim 5$.  The inferred stellar mass, $1.4\ee 4 \,\msol$ (Paumard et
al.\ 2006), places a lower limit on the disk mass: the disruptive
effect of stellar winds and radiation from the first massive stars to
form after the capture event, as well as potential losses of stars by
scattering events suggest that the initial disk mass would likely have
been at least several times higher, although SPH simulations indicate
that the process of star formation may instead be nearly 100\%
efficient (Nayakshin, Cuadra \& Springel 2007).  Thus we consider
progenitor disk masses in the range $10^4$--$10^5\,\msol$, corresponding
to $\kappa N_{24}/v_{100}^4 \sim 2\textrm{--}20$.  As noted earlier,
the theoretical value of $\lambda$ is uncertain, but likely incident
cloud speeds are between 50 and 100 \kms, so the observed disk size
implies that $\lambda\sim 0.1$--0.2.  Then the range of initial disk
masses requires cloud column densities in the range $(2-60)\ee 24 \ut
cm -2 $.  This is consistent with the observed range of
column densities of the clouds currently in the Galactic center $\sim
10^{24}-10^{25}\ut cm -2 $.

\section{Discussion}

The scenario that we have outlined leads naturally to the formation of
gravitationally-unstable rings of gas with the correct size and mass
to explain the observed stellar disk.  The mass estimate is robust as
long as the capture radius $b_0$ is smaller than the size scale of the
cloud, depending only on the idea that material that suffers
significant deflections in the central potential will collide, shock
and radiate away sufficient kinetic energy to become bound.  Indeed,
shocked, dense molecular gas cools efficiently for the $\la 100$ \kms\
shock speeds expected during circularization of the captured gas
(e.g.\ Draine, Roberge \& Dalgarno 1983; Hollenbach \& McKee 1989),
and within a few hundred years at most the temperature drops to
$\sim$\,100\,K. This is shorter than the dynamical time scale $\sim
240\,(r/0.1\un pc )^{3/2}\un yr $.  This yields captured masses
$\sim10^5\msol$ for cloud column densities $N_H\sim10^{24}\ut cm -2 $
and speeds $\sim 100$ \kms.  The size of the resultant disk is set by
the maximum angular momentum of the captured material after
circularization (represented by the parameter $\lambda$).  Although
the values of $\lambda\sim0.2$ needed to match the size of the Sgr A*
stellar disks are reasonable, fluid-dynamical simulations are
necessary to confirm this.  The estimated $Q$ value of the pre-stellar
disk is in the range 0.1--1, implying that the disk is gravitationally
unstable and should fragment once the ``turbulent'' velocity
dispersion of the gas settles down to the point that the effective Q
(with $\Delta v$ substituted for the sound speed) becomes of order
unity.  As the cooling time is comparable to the dynamical time, stars
are formed as the disk is circularizing, with a corresponding range of
eccentricities and inclinations of the orbits (cf.\ Mapelli et al.\
2008).  This mechanism may explain why the observed stellar disk(s)
are more disordered than would be produced by star formation in an
intially kinematically cold disk.  Subsequent orbit evolution due to
gravitational interactions between stars should be minimal given that
the stellar ages $\sim 6$\,Myr (Paumard et al.\ 2006) are only a few
hundred orbital periods (cf.\ Cuadra et al.\ 2008), and the resonant
relaxation timescale exceeds 30\,Myr (G\"urkan \& Hopman 2007).

\begin{figure}[tbp]
    \epsscale{0.7}
    \plotone{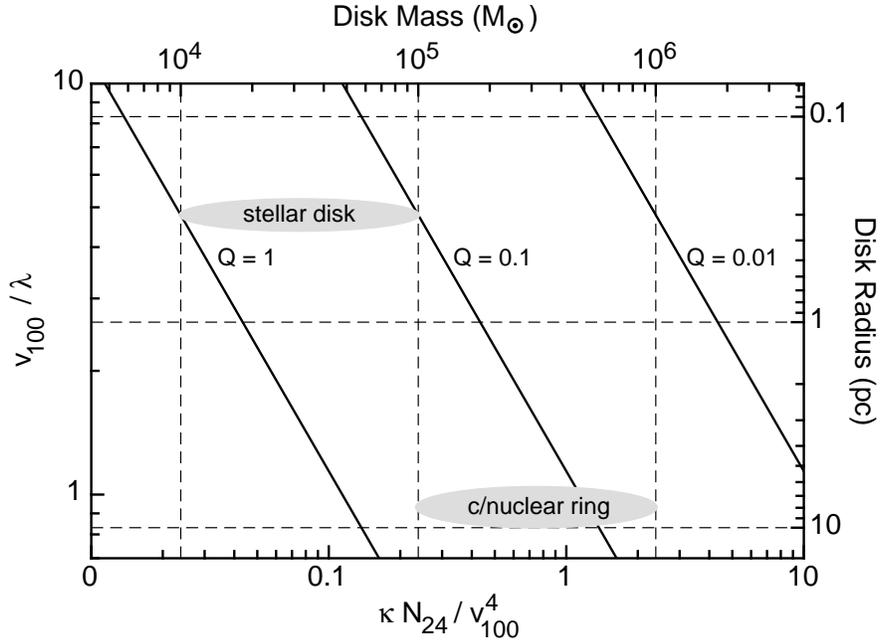}
    \caption{The mass and radius of disks
    formed by the partial capture of interloper clouds that
    temporarily engulf Sgr A* are determined by the cloud speed $v =
    v_{100}\times 100\,$\kms\ and column density
    $N_\mathrm{H}=N_{24}\times 10^{24}\ut cm -2 $.  $\kappa$ is the
    fraction of the cloud material with impact parameters less than
    $2GM/v^2$ that is captured, and $\lambda$ is the fraction of
    angular momentum remaining after circularization of the captured
    material.  Horizontal and vertical dashes indicate lines of 
    constant disk radius and mass, respectively. Diagonal lines are labelled
    by their value of $Q=c_s\Omega/\pi G \Sigma_d$ assuming a 
    temperature of 100\,K (see text); the region to the
    right of the Q=1 line are unstable to gravitational fragmentation.
    Grey shaded regions indicate the disk parameters for the the stellar disk close to Sgr A*
    and for the circumnuclear ring.}
    \label{fig:parameter_space}
\end{figure}

Turning now to the circumnuclear ring, the observed CO (Harris et al.\
1995) and HCN (G\"usten et al.\ 1987) emission indicates an outer
radius of $\sim10$\,pc and a total mass of $\sim 10^5 \msol$.  More
recent HCN observations imply that there is high density material
close to the inner edge (Jackson et al.  1993; Christopher et al.
2005) suggesting that the mass may be closer to $10^6\msol$.  From
Fig.\ \ref{fig:parameter_space} we infer $v_{100}/\lambda \approx 1$
and $\kappa N_{24}/v_{100}^4 \approx 0.3-3$.  The extent of the ring
suggests an initial cloud speed towards the lower end of the
50--100\kms\ range -- otherwise $\lambda \sim 1$, and it would not be
bound to Sgr A*.  If we adopt $v \approx 50$\kms, this implies
$\lambda \sim 0.5$.  This is reasonable given that at 50\kms\ the
Hoyle-Lyttleton radius is about 12\, pc (see eq [\ref{eq:b0}]), not
much less than the probable cloud size.  On this scale, one expects
considerable asymmetry in the cloud material passing by Sgr A* during
the capture event, with a corresponding reduction in the net
cancellation of angular momentum during the capture process (cf.\
Bottema \& Sanders 1986; Sanders 1998).  The disk mass requires column
densities in the range $(50-500)\ee 24 \ut cm -2 $, an order of
magnitude higher than typical clouds in the Galactic center region.
Note however that we ignored the gravitational effects of the evolved
stellar cluster which become important beyond 2 pc (Genzel et al.\
2000).  This tends to increase the capture radius and captured mass,
requiring larger incident cloud velocities, smaller $\lambda$ and
smaller cloud column densities.  At first sight our model suggests
that the circumnuclear ring should be severely unstable to
gravitational fragmentation, but this assumes that it has
kinematically relaxed.  The velocity dispersion of the ring is $\sim
30$\kms, so that it is not unstable unless its mass is $\ga
10^6\,\msol$.  There is little obvious sign of star formation, although
methanol and water masers -- signatures of the early phases of massive 
star formation --  have recently been detected (Yusef-Zadeh et
al.\ 2008).  It appears that the circumnuclear ring is still in the
process of settling down soon after formation.  The ring's orbital
time scale at 2 pc is $\sim 10^5$\,yr, so this implies that the age of
the ring is $\la 10^6$\,yr.  If this is so, the remains of the
original interloper cloud should lie within $\sim 100$\,pc of Sgr A*.
One candidate is the +50 \kms\ molecular cloud which extends along the
plane from the Galactic center to l$\approx0.2^0$ and consists of a
number of bound cloudlets with a total mass of $\sim10^6\,\msol$ 
(Armstrong and Barrett 1985).  This cloud is thought to lie about
30\,pc behind Sgr A*, consistent with an interaction $\sim3\ee 5 $
years ago.

The age of the stellar disk, $\sim 5\ee 6 \un yr $ (Paumard et al.\
2006), and the relative youth of the circumnuclear ring imply that the
rate of encounters of massive clouds with Sgr A* is $\sim 10^{-6}\ut
yr -1 $.  This may have been been ongoing for a significant
fraction fo the Galaxy's lifetime as the stellar population in the
central parsec is consistent with roughly constant star formation over
the past 12\,Gyr (Maness et al.\ 2007; but see also Blum et al.\
2003), The inner 200\,pc of the Galaxy is rich in dense molecular
clouds, many of which are on eccentric orbits (Bally et al.\ 1988; Oka
et al.\ 1998; Martin et al.\ 2004).  In addition to the +50 \kms\
molecular cloud, the well-known 40, 20 and -30 \kms\ molecular clouds
are all members of a disk population of molecular clouds distributed
within the inner 30\,pc of Galactic center.  Their non-circular,
elongated motion is thought to be induced by the Galaxy's barred
potential (e.g., Binney et al.\ 1991; Morris \& Serabyn 1996, and
references therein), with dynamical friction aiding migration to the
central regions of the Galaxy (Stark et al.\ 1991).  Here star
formation may instead occur through collisions between clouds, which
create a high pressure environment suitable for cluster star formation
(Tan \& McKee 2002).  For example, the intense star formation
apparent in Sgr B2 may have been triggered by the collision between
the 65 and 80 \kms molecular clouds (Mehringer et al.\  1993; Hasagawa
et al.\ 1994), and the large proper motion of the Arches cluster may
reflect this formation mechanism (Stolte et al.\ 2007).  Apart from
contributing to the central cusp in stellar density (Serabyn \& Morris
1996), the estimated infall rate, $\sim 0.4 \msol \ut yr -1 $, is more
than enough to bring a $\sim 10^5-10^6 \msol $ cloud into the inner
few parsec every few million years, where interaction with Sgr A* may
produce a burst of star formation in a sub-parsec scale stellar disk.

\clearpage

\begin{figure}[tbp]
    \epsscale{0.6}
    \plotone{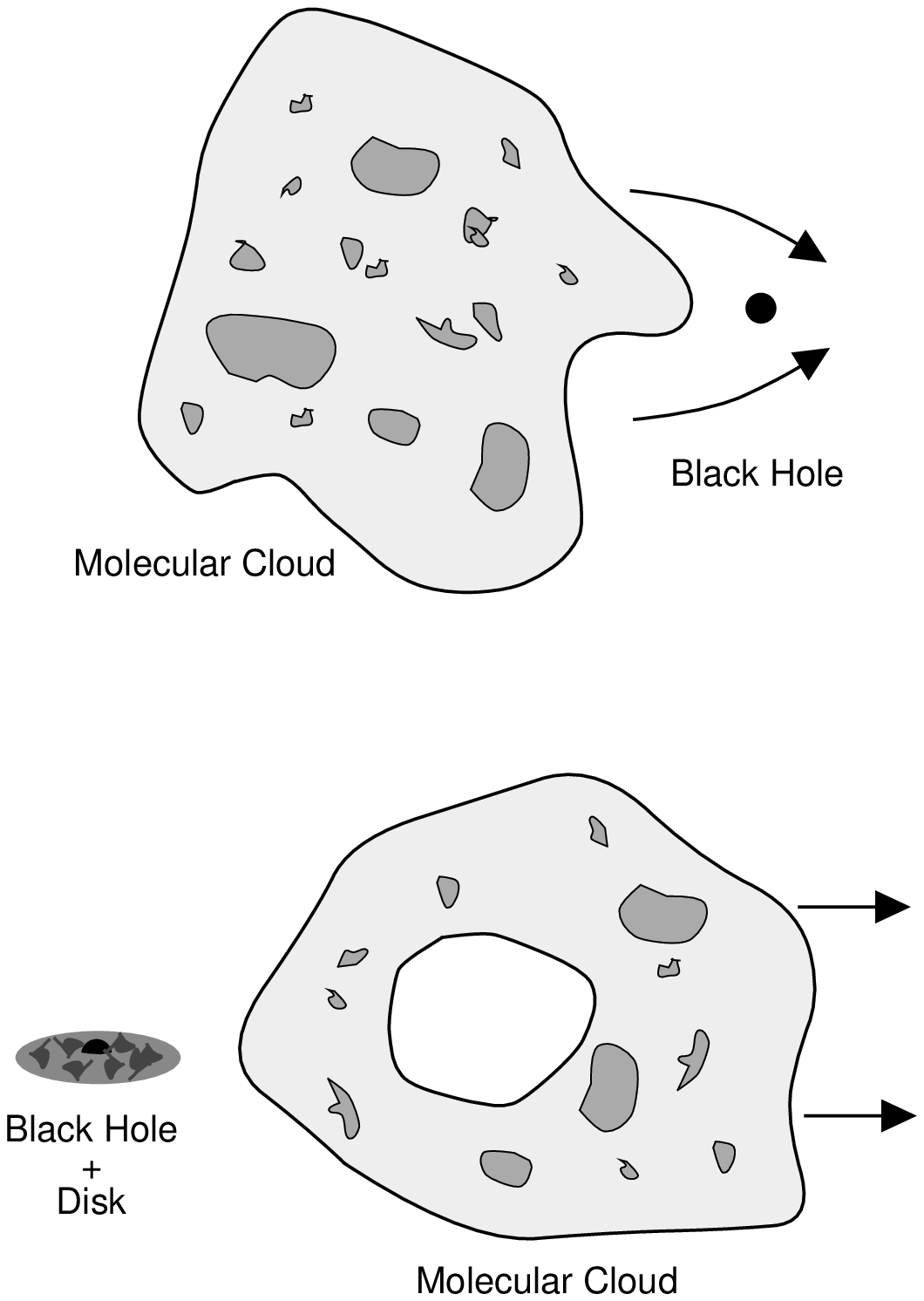}
    \caption{A schematic diagram of a cloud impacting Sgr A*.  The
    upper panel indicates the gravitational focusing of incoming
    molecular cloud material (incident from the left).  The lower
    panel shows the carved-out inner region of the cloud that has been
    captured by Sgr A* and circularized to form a disk.  The outer
    region of the cloud continues its motion in the direction away
    from Sgr A*.}
    \label{fig:cartoon}
\end{figure}

\end{document}